\documentclass[11pt]{article}
\usepackage[margin=1in]{geometry}
\usepackage{amsmath,amssymb}
\usepackage{booktabs}
\usepackage{graphicx}
\usepackage{microtype}
\usepackage{xspace}
\usepackage{url}
\usepackage[hidelinks]{hyperref}

\newcommand{\KSP}{\texttt{KSPSolve}\xspace}
\newcommand{\PtAP}{\texttt{PtAP}\xspace}

\newcommand{\MATBAIJ}{\texttt{MATBAIJKOKKOS}\xspace}

\title{A Natively Blocked, Device-Resident Algebraic Multigrid GPU Path in PETSc}

\author{Mark F.\ Adams\\
Lawrence Berkeley National Laboratory\\
Berkeley, CA 94720\\
\texttt{mfadams@lbl.gov}}
\date{}

\begin{document}
\maketitle

\begin{abstract}
Smoothed-aggregation algebraic multigrid (AMG) is widely used for the linear systems that arise from
finite-element discretizations of vector partial differential equations such as elasticity, but its
GPU implementations have used scalar sparse matrix formats. These problems carry a natural block
structure: the matrix nonzeros occur in dense $bs \times bs$ blocks that share one column index, so
storing the blocks directly removes most of the index data and raises the arithmetic intensity of
the bandwidth-bound kernels that dominate AMG on the GPU\@. Existing blocked GPU kernels in NVIDIA
cuSPARSE~\cite{cuSPARSE} and Kokkos Kernels~\cite{Rajamanickam2021} require equal row and column
block sizes, but AMG for elasticity is rectangular-blocked: the near-null space of rigid-body modes
makes the coarse block size (6 in three dimensions) differ from the fine (3), so the prolongator and
the Galerkin triple product mix block sizes. We add a portable, Kokkos-backed blocked matrix type to
PETSc~\cite{PETSc} (\MATBAIJ) with rectangular-block kernels, and make every step of the
smoothed-aggregation setup operate on the block format directly, with no expansion to scalar
\texttt{AIJ} on the coarsening path. The two phases that recur when the hierarchy is reused across
solves---the Galerkin coarse-operator recompute ($A_c = P^T A P$) and the V-cycle---are kept resident
on the device in blocks, using a native blocked off-process prolongator gather over a dedicated
\texttt{PetscSF} and a new blocked COO assembly path (\texttt{MatCOOUseBlockIndices}) that generalizes
PETSc's coordinate-format assembly to dense $bs_r \times bs_c$ blocks. On NVIDIA A100 GPUs for 3D
linear elasticity the vendor cuSPARSE Galerkin product runs out of GPU memory on a $128^3$ grid
(6.3M unknowns) packed onto 8 GPUs, where the blocked format fits; the portable native-Kokkos-Kernels
scalar path also fits there, but with a much heavier Galerkin product. Where the formats run, the
blocked format is at parity on one GPU and faster at scale: at 27 GPUs it is $1.24\times$ faster on
the V-cycle (\KSP), $1.42\times$ faster on SpMV, and $1.80\times$ faster on the coarse-operator
recompute (\PtAP), with the \PtAP advantage reaching $2.27\times$ at 64 GPUs.
\end{abstract}

\section{Introduction}
\label{sec:intro}

Multigrid methods solve the linear systems arising from elliptic PDEs in operations proportional to
the number of unknowns~\cite{Adams2010}. For unstructured, matrix-defined problems, \emph{algebraic}
multigrid (AMG) builds the multilevel hierarchy from the system matrix itself, and
\emph{smoothed-aggregation} AMG~\cite{Vanek1996} is particularly effective for systems of PDEs such
as elasticity, where the near-null space of rigid-body modes must be preserved on every coarse level.

Two hardware trends shape any modern multigrid implementation. First, accelerators (GPUs) supply the
bulk of floating-point throughput on current leadership machines, so the solver must run on the
device. Second, on those devices floating-point throughput vastly exceeds memory bandwidth, so the
cost of nearly every sparse kernel---sparse matrix--vector products in the smoother and Krylov
accelerator, sparse matrix--matrix products in the Galerkin coarse-operator construction---is set by
how many bytes move, not how many flops execute~\cite{Adams2010}.

These two trends point directly at \emph{blocked} sparse storage. A discretization of a vector PDE
associates several unknowns with each mesh point---three displacement components per node for 3D
elasticity---so the matrix has a natural fixed block size $bs$, and its nonzeros occur in dense
$bs \times bs$ blocks that share one $(row\text{-}block,\ col\text{-}block)$ index. Storing those
blocks directly amortizes one integer index over $bs^2$ values, shrinks index storage by a factor of
about $bs^2$, and raises arithmetic intensity by reusing each loaded index across an entire block.
PETSc has provided such a format (\texttt{BAIJ}) on the CPU for decades.

\textbf{Why a vendor blocked kernel is not enough.} The blocked formats in NVIDIA
cuSPARSE~\cite{cuSPARSE} and Kokkos Kernels~\cite{Rajamanickam2021} (block compressed sparse row,
BSR) require equal row and column block sizes. Smoothed-aggregation AMG for elasticity violates this
at every level transition: the near-null space---six rigid-body modes in 3D---gives each aggregate
six coarse unknowns while fine nodes carry three, so the prolongator $P$ has rectangular $3 \times 6$
blocks and the Galerkin triple product $P^T A P$ composes blocks of differing sizes. Among the vendor
libraries only AMD's rocSPARSE~\cite{rocSPARSE} exposes a general block format with independent row
and column block sizes. To obtain a portable, vendor-neutral blocked AMG we add a new
\textbf{Kokkos-backed blocked matrix type} (\MATBAIJ) whose kernels are templated on independent
row and column block sizes.

\textbf{The gap in PETSc's AMG\@.} Adding the matrix type was necessary but not sufficient. PETSc's
smoothed-aggregation multigrid (GAMG) has no blocked path: every step of its setup phase is written
against scalar \texttt{AIJ} machinery---operator inspection, the aggregation graph, the off-process
gather of prolongator rows, and the assembly of the Galerkin coarse operator all require the operator
in scalar form. A blocked operator therefore has no path through GAMG without expanding to scalar
\texttt{AIJ}, which forfeits the blocked memory advantage and, as Section~\ref{sec:capacity} shows,
can exhaust GPU memory forming the coarse Galerkin product with the vendor cuSPARSE backend.

\textbf{Contributions.} This paper makes the blocked path native end to end:
\begin{enumerate}
\item A portable, rectangular-block GPU matrix type (\MATBAIJ) with Kokkos-backed kernels supporting
  independent row and column block sizes.
\item A conversion-free blocked GAMG setup: operator inspection, the aggregation graph, the
  off-process prolongator gather, and the prolongator smoother all operate on the block format
  directly, with no \texttt{AIJ} expansion on the coarsening path.
\item Device-resident reuse of the coarse-operator recompute: the prolongator-side data is cached and
  state-gated so the hot \PtAP numeric phase incurs no host round trip on reuse.
\item A reusable blocked COO assembly primitive (\texttt{MatCOOUseBlockIndices}) generalizing
  PETSc's coordinate-format assembly to dense $bs_r \times bs_c$ blocks, of independent value for
  finite-element assembly.
\item A measured characterization of when blocked wins, with an arithmetic-intensity argument for
  why the dominant kernels are bandwidth-bound.
\end{enumerate}

The remainder is organized as follows. Section~\ref{sec:background} reviews multigrid and block
structure. Section~\ref{sec:design} describes the native blocked design.
Section~\ref{sec:performance} reports performance on A100 GPUs. Section~\ref{sec:coo} presents the
blocked COO primitive. Section~\ref{sec:conclusion} concludes.

\section{Background}
\label{sec:background}

\subsection{Multigrid methods}

Multigrid is motivated by the observation that a low-resolution discretization of an operator can
capture error components that are expensive to resolve on a highly resolved discretization. A simple
pointwise relaxation (smoother) such as Gauss--Seidel or damped Jacobi efficiently damps the
high-frequency, locally determined part of the error but leaves the smooth, globally coupled part
nearly untouched; a coarser representation resolves that smooth error cheaply. Applying this idea
recursively over a sequence of coarser levels yields a method whose total work converges to a small
multiple of the finest-level work.

For a linear system $Ax = b$, a two-level cycle is defined by a smoother $S$, restriction operator
$R$, prolongation operator $P$, and coarse operator $A_c$. One cycle pre-smooths the iterate, forms
and restricts the residual, solves the coarse correction, prolongs and adds it, and post-smooths.
Recursion gives the V-cycle. The coarse operator is most commonly formed by the Galerkin triple
product $A_c = P^T A P$. A consideration growing steadily more important is that modern processors
have limited memory bandwidth relative to floating-point throughput, and minimizing memory traffic is
therefore a primary design criterion~\cite{Adams2010}.

\subsection{Algebraic multigrid by smoothed aggregation}

\emph{Algebraic multigrid} (AMG) constructs the entire hierarchy from the system matrix. We use the
\emph{smoothed-aggregation} (SA) variant~\cite{Vanek1996}, which is well suited to systems of PDEs
and to the block structure exploited here.

SA builds each coarse level in two stages: aggregation and prolongator construction, driven by
\emph{strong coupling}. For threshold $\varepsilon$, the strongly-coupled neighborhood of node $i$ is
\[
  N_i(\varepsilon) = \{ j : |a_{ij}| \ge \varepsilon \sqrt{a_{ii}\, a_{jj}} \}.
\]
Aggregates are formed by a greedy disjoint covering. Each aggregate gives rise to one coarse node. A
\emph{tentative} prolongator $\tilde{P}$ is defined so that the near-null space is represented
exactly on the coarse level. For 3D elasticity the near-null space is the six rigid-body modes, so
each aggregate contributes six coarse degrees of freedom and the coarse operator naturally carries a
larger block size than the fine operator (e.g.\ $bs$: 3 $\to$ 6). Preserving these zero-energy modes
on every level is essential for mesh-independent convergence~\cite{Vanek1996}.

The tentative prolongator is improved by one step of a damped-Jacobi \emph{prolongator smoother},
\[
  P = (I - \omega D^{-1} A)\tilde{P},
\]
where $D$ is the diagonal of $A$. The coarse operator is then $A_c = P^T A P$ and the process
repeats.

\subsection{Block structure of systems of PDEs}

Discretizations of vector PDEs associate several unknowns with each mesh point---three displacement
components per node for 3D elasticity---so the system matrix has a natural fixed block size $bs$, and
its nonzeros occur in dense $bs \times bs$ blocks that share a single $(row\text{-}block,\
col\text{-}block)$ index. Storing and operating on these blocks directly, rather than as $bs^2$
independent scalar entries each carrying its own column index, is the classical motivation for
PETSc's \texttt{BAIJ} format.

Crucially, the block structure is \emph{not uniform across levels}: the fine elasticity operator has
$3 \times 3$ blocks, but the coarse block size equals the dimension of the preserved near-null
space---six rigid-body modes in 3D---so coarse operators carry $6 \times 6$ blocks and the
prolongator has rectangular $3 \times 6$ blocks. A blocked AMG hierarchy therefore demands kernels
that handle independent row and column block sizes.

\subsection{The blocked-GPU gap in PETSc AMG}

PETSc's GPU support routes sparse kernels through backend libraries: cuSPARSE~\cite{cuSPARSE} for
scalar \texttt{AIJ} on NVIDIA hardware, and Kokkos Kernels~\cite{Rajamanickam2021} for the portable
\texttt{*KOKKOS} types. Scalar AMG on the GPU is mature. The vendor libraries do provide blocked
sparse formats, but both cuSPARSE and Kokkos Kernels implement BSR with a single fixed square $bs$,
which cannot represent the rectangular $3 \times 6$ prolongator or mixed-size Galerkin products of
Section~\ref{sec:background}. Among the major vendor libraries only AMD's rocSPARSE~\cite{rocSPARSE}
offers a general block format (GEBSR) with independent row and column block sizes. To obtain a
portable blocked AMG we implement a Kokkos-backed blocked matrix type with rectangular-block kernels.

Even with that matrix type in hand, PETSc has no device-resident blocked GAMG setup: the GAMG setup
is written against scalar \texttt{AIJ}, so operator inspection, the aggregation graph, and coarse-operator
assembly all require the operator in scalar form, making a conversion-free blocked path impossible without
this work.

\section{Design}
\label{sec:design}

The design rests on two invariants. First, \textbf{the blocked operator is never expanded to scalar
\texttt{AIJ} anywhere on the coarsening path}. Second, \textbf{the two recurring phases of a
production solve---the Galerkin coarse-operator recompute (\PtAP) and the V-cycle---are fully
device-resident in blocks}. The one-time aggregation graph setup remains a host computation
(Section~\ref{sec:graph}); because the hierarchy is reused, it is a cold, amortized cost outside the
recurring work.

\subsection{What is hot, and why it matters}
\label{sec:hot}

A GAMG solve has two phases. The \emph{setup} coarsens the operator: it builds the aggregation graph,
the tentative and smoothed prolongators, and the Galerkin coarse operators $A_c = P^T A P$ on every
level. The \emph{solve} runs the V-cycle. In a production nonlinear or time-dependent simulation the
operator $A$ changes on every Newton or time step, but the hierarchy---the aggregates and
prolongators---is reused across steps (\texttt{-pc\_gamg\_reuse\_interpolation}). On each step the
\emph{symbolic} setup is computed once and reused, while the \emph{numeric} coarse operators are
recomputed because $A$ is new. The recurring operations are therefore:
\begin{itemize}
\item \textbf{hot \PtAP}: the numeric Galerkin recompute $A_c = P^T A P$ with reused sparsity
  (\texttt{MAT\_REUSE\_MATRIX}), on every level; and
\item \textbf{hot \KSP}: the V-cycle, dominated by SpMV in the smoother and Krylov accelerator.
\end{itemize}

\subsection{Native block operator inspection and aggregation graph (host)}
\label{sec:graph}

Smoothed aggregation needs two pieces of information from the operator before forming any product: a
scalar measure of each block row for the strength-of-connection threshold, and a graph whose edges
are the strong couplings $N_i(\varepsilon)$. GAMG's existing code requires a scalar \texttt{AIJ}
operator to supply them.

We compute both directly from the block format. Operator inspection
(\texttt{MatGetInfo}/\texttt{MatNorm}) is computed over the $bs \times bs$ blocks of the Kokkos
\texttt{BAIJ} storage. The aggregation graph is built from the block sparsity: each $(row\text{-}block,\
col\text{-}block)$ index is one graph edge, and the strength weight is the block norm. These
operations run on the \textbf{host}---graph construction is irregular, serial-leaning work poorly
suited to the GPU---but, critically, they no longer pay the $bs^2$ scalar expansion, and because the
graph and aggregates are built once and reused across solves they sit in the cold setup outside the
recurring per-step cost.

\subsection{Native block off-process prolongator gather ($P_{\mathrm{oth}}$)}
\label{sec:poth}

The Galerkin product $P^T A P$ on a distributed matrix needs, on each rank, the rows of $P$
corresponding to the off-process columns of the local $A$---the off-process prolongator rows
$P_{\mathrm{oth}}$. We carry the gather natively in the blocked format. A dedicated \texttt{PetscSF}
is constructed whose roots and leaves are \emph{block rows} of $P$, and the broadcast moves whole
prolongator blocks (\texttt{PetscSFBcast} over a block-typed unit) rather than separate scalars.
Because $P$ is rectangular ($3 \times 6$ blocks), the block-typed unit is the full $bs_r \times
bs_c$ payload. The communication is GPU-aware: the block payload is staged and exchanged directly
from device memory (\texttt{PetscSFBcastWithMemType} with \texttt{PETSC\_MEMTYPE\_KOKKOS}), so the
gather never round-trips through the host.

\subsection{Blocked COO assembly of the coarse operator}
\label{sec:coo-design}

The coarse operator is assembled from local triple-product contributions plus off-process
contributions routed to their owning ranks. PETSc's device-assembly path uses \emph{coordinate
format} (COO): the caller supplies, once, the $(i, j)$ coordinates of every contribution
(\texttt{MatSetPreallocationCOO}), PETSc builds a cached communication-and-scatter plan, and
thereafter each numeric assembly is a single \texttt{MatSetValuesCOO} that scatters a flat value
array and sums duplicates, entirely on the device.

We generalize this to blocks. A new option, \texttt{MatCOOUseBlockIndices}, tells the COO setup that
each coordinate addresses a dense $bs_r \times bs_c$ block. The preallocation plan is built over
block coordinates, and the numeric scatter moves block payloads. The block size need not be
square---the coarse-operator assembly uses the same mixed sizes as the Galerkin product---so the
primitive covers the rectangular case the vendor square-block formats cannot.

\subsection{Device-resident reuse of the hot \PtAP}
\label{sec:reuse}

The three pieces above make a \emph{single} blocked \PtAP native. The production case repeats it
under reused interpolation, where $P$ is \emph{fixed} across hot solves and only $A$ changes. We
exploit this by caching the prolongator-side data---the local prolongator rows $P_{\mathrm{local}}$,
the off-process rows $P_{\mathrm{oth}}$, the stacked operand, and the transpose $R$---and gating its
rebuild on the prolongator's object state (\texttt{PetscObjectState}). On a hot recompute, if $P$'s
state matches the cached value, the path reuses the cached device-resident values directly: in
particular the $P_{\mathrm{oth}}$ gather is not re-broadcast but served from cache. The hot \PtAP
numeric phase becomes a local blocked triple product plus the off-process reduction of the new coarse
values, with everything on the prolongator side served from device-resident cache.

Combined, Sections~\ref{sec:graph}--\ref{sec:reuse} deliver the two invariants. No step of the
coarsening path expands the operator to scalar \texttt{AIJ}, and across the recurring phases the
operator stays in blocks on the device with no host round trip on reuse.

\section{Performance}
\label{sec:performance}

\subsection{Experimental setup}
\label{sec:setup}

All results are on the Perlmutter system at NERSC\@. Each node has four NVIDIA A100 GPUs, and we use
GPU-aware MPI. Unless noted we run \textbf{one MPI rank per A100}: sharing a GPU between two ranks
inflated contention and is not representative of production placement. The build is optimized, with
the CUDA backend for Kokkos~\cite{Trott2022} and Kokkos Kernels~\cite{Rajamanickam2021} configured
with NVIDIA's vendor libraries (cuSPARSE, cuBLAS, cuSOLVER) as third-party libraries (TPLs)---the
default for a CUDA \texttt{--download-kokkos-kernels} build. With these TPLs enabled, the scalar
\texttt{AIJKOKKOS} matrix type dispatches its sparse kernels to \textbf{cuSPARSE}; built without
them the same type runs the portable native \textbf{Kokkos Kernels (KK)} implementations. The
blocked kernels are our own in either build: cuSPARSE's BSR format supports only square, equal-size
blocks, so rectangular Galerkin and prolongator products cannot route through it. We use
\emph{cuSPARSE} and \emph{Kokkos Kernels (KK)} to name the two scalar backends, reserving \emph{TPL}
for the build-configuration discussion (Section~\ref{sec:backends}).

The model problem is 3D linear elasticity (block size $bs = 3$), preconditioned by GAMG with a
point-block Jacobi smoother (\texttt{pbjacobi}) and a CG Krylov accelerator.
Sections~\ref{sec:kspsmv}--\ref{sec:capacity} and~\ref{sec:bw}--\ref{sec:cold} use the
hand-assembled trilinear (Q1) hexahedral elasticity example
\texttt{src/ksp/ksp/tutorials/ex56}~\cite{PETSc}. We identify each problem by its node grid $m^3$; the
weak-scaling ladder is $m^3 \in \{32^3, 64^3, 96^3, 128^3\}$ for $\{1, 8, 27, 64\}$ GPUs, holding
98,304 unknowns per GPU\@. The separate nonzeros-per-row study of Section~\ref{sec:q1q2} uses a
DMPlex/PetscFE harness (\texttt{src/snes/tutorials/ex56}) with selectable element order (Q1/Q2).
The two harnesses use different boundary/load configurations and are not directly comparable across
tables. We use the unpreconditioned residual norm throughout; with this norm the two formats converge
in the \emph{same} iteration count to the same true residual on every problem we report.

We compare three GPU backends: two \textbf{scalar} (\texttt{AIJKOKKOS}) backends---\textbf{cuSPARSE}
(the vendor-tuned default) and \textbf{Kokkos Kernels (KK)} (TPLs off)---and one \textbf{block}
(\texttt{BAIJKOKKOS}) backend. cuSPARSE is the primary baseline. We follow the production model of
Section~\ref{sec:hot}: a hierarchy is built once and reused across solves with a changing operator
(\texttt{-pc\_gamg\_reuse\_interpolation true}, \texttt{-two\_solves}). \textbf{Hot} numbers are
taken from the second/third solve and the second setup.

\textbf{A note on timers.} PETSc's \texttt{-log\_view} mis-attributes asynchronous GPU kernel time
unless per-event device synchronization is enabled (\texttt{-log\_view\_gpu\_time}), but that
synchronization also serializes MPI waits and inflates end-to-end solve time. We read each metric
from the run in which it is accurate: \KSP from the run \emph{without} the GPU timer (true wall
time), and SpMV from the run \emph{with} it (true GPU-event time). For the communication-bound hot
\PtAP, \emph{neither} \texttt{-log\_view} mode is trustworthy, so we report hot \PtAP from Nsight
Systems traces scoped to the hot \texttt{MatPtAPNumeric} spans.

\subsection{Hot \KSP and SpMV---block wins at scale}
\label{sec:kspsmv}

\begin{figure}[t]
  \centering
  \includegraphics[width=0.72\linewidth]{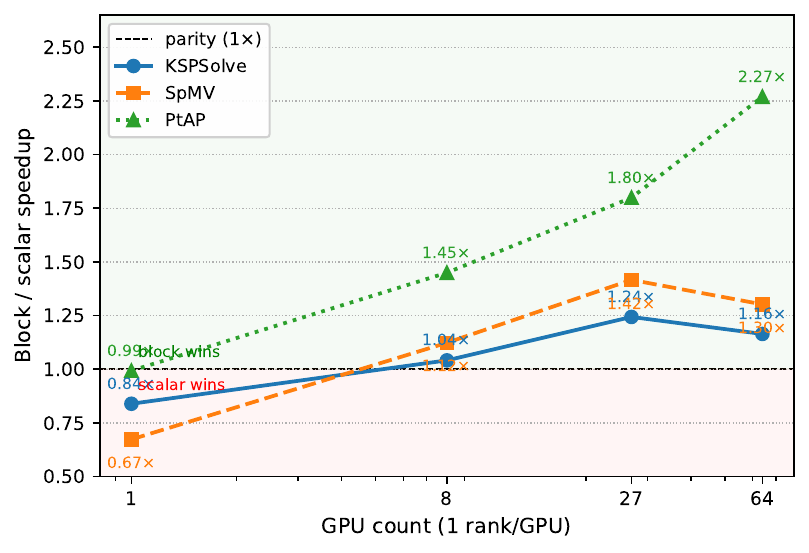}
  \caption{Block/scalar speedup ratio (scalar time / block time; $>1$ means block faster) at one
  rank per A100. SpMV advantage peaks at 27 GPUs ($1.42\times$, at the theoretical
  index-bandwidth ceiling) and narrows slightly at 64 GPUs as cuSPARSE handles the larger
  communication pattern more efficiently. \PtAP advantage grows monotonically ($1.45\times \to
  2.27\times$).}
  \label{fig:speedup}
\end{figure}

The V-cycle is SpMV-bound: the matrix--vector products in the smoother and CG accelerator account
for the large majority of \KSP time. Table~\ref{tab:weak} gives the weak-scaling comparison, and
Figure~\ref{fig:speedup} summarizes the speedup ratios across the weak-scaling ladder.

\begin{table}[t]
\caption{Hot timings (seconds), weak scaling at one rank per A100, 3D elasticity. \KSP is
off-timer wall time; SpMV is gpu-timer; \PtAP is Nsight-scoped. Iteration counts match at each
scale (1 GPU: 24; 8 GPUs: 30; 27 GPUs: 35; 64 GPUs: 38). \textbf{Bold} = faster.}
\label{tab:weak}
\small
\setlength{\tabcolsep}{4pt}
\centering
\begin{tabular}{lcccccccc}
\toprule
& \multicolumn{2}{c}{1 GPU, $32^3$}
& \multicolumn{2}{c}{8 GPUs, $64^3$}
& \multicolumn{2}{c}{27 GPUs, $96^3$}
& \multicolumn{2}{c}{64 GPUs, $128^3$} \\
\cmidrule(lr){2-3}\cmidrule(lr){4-5}\cmidrule(lr){6-7}\cmidrule(lr){8-9}
Event (hot) & Block & Scalar & Block & Scalar & Block & Scalar & Block & Scalar \\
\midrule
\KSP (s)        & 0.0510 & \textbf{0.0428} & \textbf{0.1428} & 0.1487 & \textbf{0.2718} & 0.3380 & \textbf{0.2793} & 0.3252 \\
SpMV (s)        & 0.0361 & \textbf{0.0243} & \textbf{0.0893} & 0.1003 & \textbf{0.1492} & 0.2113 & \textbf{0.1632} & 0.2125 \\
\PtAP (ms, scoped) & 5.99  & \textbf{5.95}  & \textbf{10.21} & 14.80  & \textbf{10.50} & 18.89  & \textbf{10.88} & 24.69  \\
\bottomrule
\end{tabular}
\end{table}

At eight GPUs the block format wins both \KSP (0.143 vs.\ 0.149\,s, $1.04\times$) and SpMV (0.089
vs.\ 0.100\,s, $1.12\times$). The advantage widens at 27 GPUs: \KSP 0.272 vs.\ 0.338\,s ($1.24\times$)
and SpMV 0.149 vs.\ 0.211\,s ($1.42\times$). At 64 GPUs the block format continues to win: \KSP
0.279 vs.\ 0.325\,s ($1.16\times$) and SpMV 0.163 vs.\ 0.213\,s ($1.30\times$). The block SpMV moves
one column index per $bs \times bs$ block instead of $bs^2$ indexed scalar entries, moving less data
per useful flop on a bandwidth-bound device. The SpMV advantage peaks at 27 GPUs ($1.42\times$) and
narrows slightly at 64 GPUs, where scalar cuSPARSE handles the larger-rank communication pattern more
efficiently. On a single GPU the scalar path wins \KSP/SpMV by $\sim19$--$33\%$: with no halo
communication to amortize the index traffic against, the tuned scalar cuSPARSE SpMV is faster.

\textbf{Bounding the SpMV advantage.} Counting bytes a $3 \times 3$ block (nine scalar nonzeros)
moves in each format, with \texttt{int32} indices and \texttt{fp64} values:
\begin{center}
\small
\begin{tabular}{lrrrr}
\toprule
Format & Values & Indices & Total & Per scalar nz \\
\midrule
Scalar CSR  & $9 \times 8 = 72$\,B & $9 \times 4 = 36$\,B & 108\,B & 12.0\,B \\
Blocked BAIJ & $9 \times 8 = 72$\,B & $1 \times 4 =  4$\,B &  76\,B &  8.44\,B \\
\bottomrule
\end{tabular}
\end{center}
The blocked format moves 76/108 = \textbf{0.70} of the scalar matrix bytes---a 30\% reduction, for a
traffic-bound ceiling of $108/76 \approx 1.42\times$. The reason is that \texttt{fp64} values dominate:
the index is only 4 of the 12\,B per scalar nonzero, and blocking removes 8 of every 9 indices,
saving at most $(8/9) \cdot 4/12 \approx 30\%$. The measured $1.42\times$ at 27 GPUs is at this
theoretical ceiling, so the blocked SpMV realizes essentially all of the available index-bandwidth
saving. The same accounting explains why the largest effect is on \emph{capacity}
(Section~\ref{sec:capacity}), not SpMV speed: the scalar out-of-memory failure comes from the
$bs^2$-expanded SpGEMM symbolic buffers in the Galerkin product, a far larger factor than the 30\%
steady-state SpMV saving.

\subsection{cuSPARSE vs.\ Kokkos Kernels (scalar backends)}
\label{sec:backends}

\begin{figure}[t]
  \centering
  \includegraphics[width=0.72\linewidth]{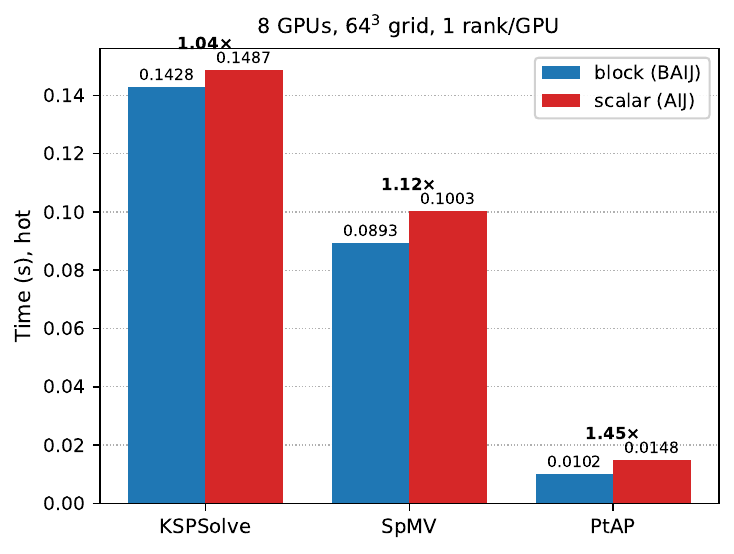}
  \caption{Hot timings at 8 GPUs ($64^3$ grid, one rank per A100). Grouped bars show block (dark)
  vs.\ scalar cuSPARSE (light) for \KSP, SpMV, and \PtAP\@. \PtAP values are Nsight-scoped.}
  \label{fig:n8}
\end{figure}

To verify that the block win is measured against the stronger scalar backend, we rebuilt from the
same source with \texttt{--with-kokkos-kernels-tpl=0}, so the scalar \texttt{AIJKOKKOS} type runs
native Kokkos Kernels (KK) instead of cuSPARSE. The TPL choice affects only the scalar backend;
the block kernels are the same code in both builds. Table~\ref{tab:backends} shows the result.

\begin{table}[t]
\caption{Scalar backend with and without the cuSPARSE TPL, Q1 $n=8$, $64^3$ grid. SpMV is
gpu-timer; \PtAP is Nsight-scoped (ms). Block (\texttt{BAIJ}) is the same code in both builds
($\pm1\%$ run-to-run variance). \textbf{Bold} = faster scalar backend.}
\label{tab:backends}
\centering
\begin{tabular}{lccc}
\toprule
Metric (hot) & Scalar cuSPARSE & Scalar KK & Block (BAIJ) \\
\midrule
SpMV (s)         & 0.1015          & \textbf{0.0949} & 0.0886 \\
\KSP (s)         & 0.1496          & \textbf{0.1368} & 0.1408 \\
\PtAP (ms)       & \textbf{14.60}  & 111.95          & 9.80   \\
\bottomrule
\end{tabular}
\end{table}

At this index-bound, mid-size configuration native KK CSR SpMV is faster than cuSPARSE
(0.0949 vs.\ 0.1015\,s, $-6\%$). The block backend still wins both scalar backends on SpMV:
$1.07\times$ over KK and $1.15\times$ over cuSPARSE. Because KK is the faster scalar SpMV here, the
weak-scaling SpMV ratios in Table~\ref{tab:weak}, computed against cuSPARSE, slightly overstate the
block advantage relative to the best scalar backend. The \PtAP comparison goes the other way: native
KK \texttt{spgemm} is $7.7\times$ slower than cuSPARSE for the hot numeric phase
(112 vs.\ 14.6\,ms), so cuSPARSE is the stricter \PtAP baseline. We report cuSPARSE throughout as the
vendor-default scalar backend.

\subsection{Hot \PtAP---device-resident reuse}
\label{sec:ptap}

The Galerkin coarse-operator recompute is the second production-critical metric, and it is where the
device-resident reuse of Section~\ref{sec:reuse} has the largest effect. Table~\ref{tab:ptap} shows
the ablation at eight GPUs.

\begin{table}[t]
\caption{Hot \PtAP at 8 GPUs ($64^3$ grid), Nsight-scoped, milliseconds. ``Ungated'' re-broadcasts
$P_{\mathrm{oth}}$ every recompute; ``state-gated'' reuses it from cache when $P$ is unchanged.}
\label{tab:ptap}
\centering
\begin{tabular}{lccc}
\toprule
Inside hot \PtAP & Block ungated & Block state-gated & Scalar (cuSPARSE) \\
\midrule
Triple-product compute & 7.83 & 7.4  & 10.57 \\
$P_{\mathrm{oth}}$ broadcast & 9.93 & \textbf{0} & 3.17 \\
Off-process reduce & 13.94 & 2.2 & 0.45 \\
\midrule
\textbf{Wall total} & 31.85 & \textbf{10.2} & 14.80 \\
\bottomrule
\end{tabular}
\end{table}

Without device-resident reuse, the block hot \PtAP at eight GPUs is 31.8\,ms (Table~\ref{tab:ptap},
ungated), against the scalar path's 14.8\,ms---$2.15\times$ slower. Nsight scoping locates the gap in
communication, not arithmetic: the block triple-product compute (7.83\,ms) is faster than cuSPARSE's
\texttt{spgemm} (10.57\,ms), but the ungated path repeats the $P_{\mathrm{oth}}$ \texttt{PetscSF}
broadcast and the off-process reduction inside every numeric recompute (together $\sim24$\,ms), work
that the scalar cuSPARSE path folds into its cached symbolic phase.

State-gating the prolongator-side rebuild removes this. With $P$ unchanged the cached $P_{\mathrm{oth}}$
gather is reused rather than re-broadcast, so its cost falls to zero and the device-to-host transfers
fall to zero; removing the redundant per-rank work also tightens the coarse-grid synchronization,
shrinking the imbalance-gated wait at the off-process reduction from 13.9 to $\sim2.2$\,ms. The block
hot \PtAP drops from 31.8 to 10.2\,ms ($3.1\times$), moving from $2.15\times$ slower than scalar to
$1.45\times$ faster. On a single GPU there is no communication and the \PtAP comparison is pure
compute: block 5.99\,ms vs.\ scalar 5.95\,ms, within 1\% (parity).

\subsection{Memory capacity---a backend-specific limit}
\label{sec:capacity}

\begin{figure}[t]
  \centering
  \includegraphics[width=0.72\linewidth]{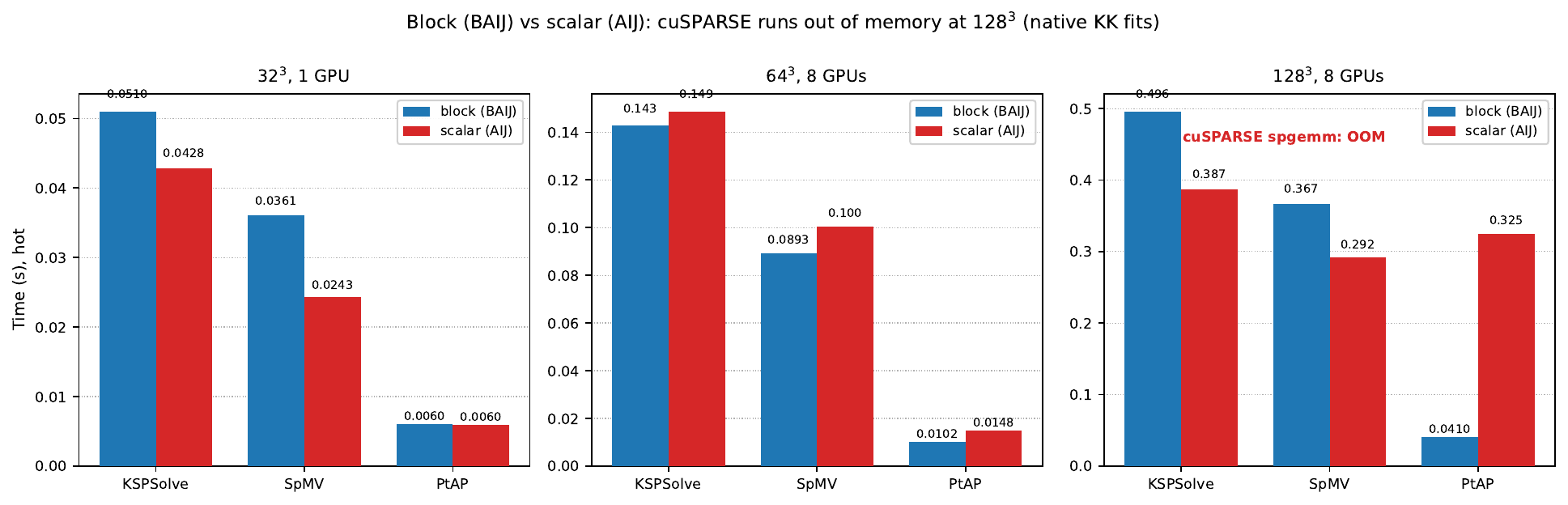}
  \caption{Capacity at $128^3$ on 8 A100s ($8\times$ the weak-scaling per-GPU load). The scalar
  \texttt{AIJ} path backed by cuSPARSE runs out of memory forming the first coarse Galerkin product;
  the portable native Kokkos Kernels backend (shown) and the blocked format both fit, in 41
  iterations. The $32^3$ and $64^3$ panels (scalar = cuSPARSE) are for reference. Hot timings: \KSP
  off-timer, SpMV/\PtAP gpu-timer. At $128^3$ the blocked Galerkin product (\PtAP) is $\sim 8\times$
  cheaper than native-KK scalar, while the compute-bound V-cycle slightly favors scalar.}
  \label{fig:oom}
\end{figure}

A memory effect appears at the largest local problem sizes, and it is backend-specific. The scalar
GPU format expands every $bs \times bs$ block into $bs^2$ scalar entries, each with its own column
index, so its sparse matrix--matrix product allocates much larger temporary buffers than the blocked
equivalent. Placing a $128^3$ grid (6.3M unknowns) on only eight A100 GPUs---eight times the per-GPU
load of the weak-scaling ladder---the scalar \texttt{AIJ} path backed by cuSPARSE runs out of GPU
memory: the \texttt{cudaMalloc} of the cuSPARSE \texttt{spgemm} symbolic buffer fails forming the
first coarse-level Galerkin product.

This is a property of the cuSPARSE \texttt{spgemm}, not of the scalar format. Rebuilt with the
portable native Kokkos Kernels \texttt{spgemm} (the TPL-off backend of Section~\ref{sec:backends}),
the same scalar \texttt{AIJ} problem fits and converges in 41 iterations, matching the blocked format.
The two paths differ sharply in the Galerkin product: the native-KK scalar coarse-operator recompute
costs $\approx 0.33$\,s and dominates its setup, against $\approx 0.04$\,s for the blocked format
(gpu-timer), the $\sim 8\times$ gap of the slow native-KK \texttt{spgemm} seen in
Section~\ref{sec:backends}. At this $8\times$-overloaded, compute-bound load the scalar V-cycle is in
fact slightly faster than the blocked one (hot \KSP 0.39 vs.\ 0.50\,s), as in the
high-nonzeros-per-row regime of Section~\ref{sec:q1q2}; the blocked advantage at this size is the
Galerkin product, not the V-cycle. Two readings follow (Figure~\ref{fig:oom}): against the
vendor-default cuSPARSE backend the blocked format extends the problem size a machine can solve;
against the portable backend it removes the Galerkin product as the dominant cost.

\subsection{Sensitivity to nonzeros per row (Q1 vs.\ Q2)}
\label{sec:q1q2}

\begin{figure}[t]
  \centering
  \includegraphics[width=0.72\linewidth]{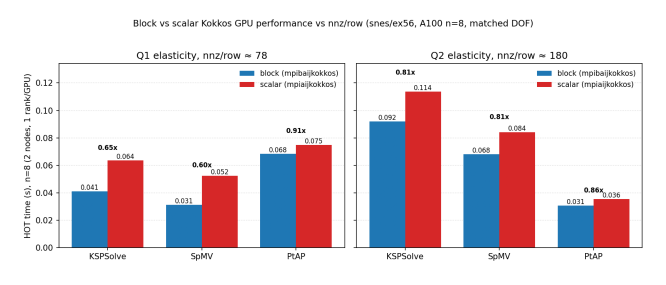}
  \caption{Block $\div$ scalar hot-timing ratios (lower = block faster) for Q1 ($\sim78$
  nonzeros/row) and Q2 ($\sim180$ nonzeros/row) at one and eight GPUs. The block advantage shrinks
  with nonzeros per row at eight GPUs, as the kernels shift from index-bound to more compute-bound.}
  \label{fig:q1q2}
\end{figure}

A natural hypothesis is that the block advantage \emph{grows} with nonzeros per row: more entries per
row amortize the block index overhead further and raise arithmetic intensity. We tested it with a
finite-element harness comparing Q1 ($\sim78$ nonzeros/row) against Q2 ($\sim180$ nonzeros/row), at
one and eight GPUs, matched in total degrees of freedom. Table~\ref{tab:q1q2} and
Figure~\ref{fig:q1q2} show the result.

\begin{table}[t]
\caption{Block $\div$ scalar ratio ($<1$ means block faster), hot solve, finite-element harness.}
\label{tab:q1q2}
\centering
\begin{tabular}{llrcc}
\toprule
Size & Degree & nnz/row & \KSP & SpMV \\
\midrule
$n=1$ & Q1 &  76 & 1.19 & 1.44 \\
$n=1$ & Q2 & 169 & 1.33 & 1.62 \\
$n=8$ & Q1 &  78 & \textbf{0.65} & \textbf{0.60} \\
$n=8$ & Q2 & 180 & \textbf{0.81} & \textbf{0.81} \\
\bottomrule
\end{tabular}
\end{table}

The hypothesis is \textbf{refuted}. At one GPU the block format loses, and the deficit \emph{grows}
with nonzeros per row (SpMV $1.44\times \to 1.62\times$ slower). At eight GPUs the block format wins,
but its advantage \emph{shrinks} with nonzeros per row (SpMV $1.67\times \to 1.23\times$ faster).
The reason is that the block format's edge comes from compressing column indices---a benefit that is
proportionally largest in the index-bound, low-nonzeros-per-row regime. As nonzeros per row grow,
the kernels become more flop-bound, where the scalar cuSPARSE implementation is strongest and the gap
closes. The low-nonzeros-per-row Q1 discretization is thus the better case for the block speed
advantage.

\subsection{Why the kernels are bandwidth-bound}
\label{sec:bw}

The performance story rests on the dominant kernels being bandwidth-bound. The simplest case is the
SpMV accounted above: the block format moves 76\,B per $3 \times 3$ block against 108\,B for scalar
CSR, and the resulting $1.42\times$ ceiling is reached at 27 GPUs. The Galerkin SpGEMM is bandwidth-
bound as well. A block-pair product moves $O(bs^2)$ bytes of $A$ and $P$ values plus one amortized
4-byte index while performing $O(bs^3)$ flops, so its arithmetic intensity grows with $bs$ but stays
far below the A100 double-precision ridge of $\approx 6.2$\,flop/byte at both $bs = 3$ and $bs = 6$;
the scalar \texttt{spgemm}, at $\approx 0.17$\,flop/byte (a 12-byte value-plus-index per 2-flop
entry), is lower still. Floating-point peak is therefore not the limiting factor---all configurations
run at $<5\%$ of fp64 peak---and data volume is.

ncu measurements on a single A100 (one rank, Q1 and Q2 harnesses) confirm both kernels are well
below the ridge. Table~\ref{tab:ncu} shows the dominant fine-level kernel for each backend.

\begin{table}[t]
\caption{ncu metrics for the dominant fine-level SpGEMM kernel, single A100. Block kernel:
\texttt{RunNumericAB\_SeqBAIJKokkos<3,3,6>}. Scalar: cuSPARSE \texttt{load\_balancing\_kernel}.
All runs unmodified GPU clocks (no clock lock).}
\label{tab:ncu}
\centering\small
\begin{tabular}{lccc}
\toprule
Config & DRAM traffic & \%peak HBM & \%peak SM \\
\midrule
Block, Q2, fine $\langle 3,3,6\rangle$ & 415\,MB & 8.9\% & 19.6\% \\
Scalar, Q2, fine                        & 4.22\,GB & 46.5\% & 22.9\% \\
Block, Q1, fine $\langle 3,3,6\rangle$ & 163\,MB &  9.4\% & 21.0\% \\
\bottomrule
\end{tabular}
\end{table}

The DRAM traffic ratio is $4{,}220\,\text{MB} / 415\,\text{MB} \approx 10.2\times$, close to the
theoretical $bs^2 = 9\times$, confirming the index-amortization argument. Both kernels sit far
below the A100 ridge ($6.2$\,flop/byte): even the scalar kernel at 46.5\% of peak HBM
($\approx 930$\,GB/s) is bandwidth-dominated. The block kernel reaches a lower fraction of peak HBM
(8.9\%, $\approx 180$\,GB/s) because the matrices are small at $n=1$, but it wins in absolute time
because it moves $10\times$ less data: the block format's advantage is a reduction in
\emph{data volume}, not a higher achieved bandwidth fraction.

\subsection{Communication profile and load imbalance}
\label{sec:comm}

The hot \PtAP at scale is gated by an exposed, load-imbalance-driven wait at the off-process
reduction. The block format moves fewer messages in every case (one block reduce versus $bs^2$ scalar
reduces per entry), so its fastest rank waits the least; the imbalance ratio is correspondingly
higher for block (e.g.\ $75\times$ at Q1, $94\times$ at Q2 versus $15\times$ and $36\times$ for scalar)
precisely because the block path's fast ranks finish their lighter communication sooner and then wait
on the slow rank. This is consistent with the scoped \PtAP story of Section~\ref{sec:ptap}: the
block path moves less, so what remains is dominated by the coarse-grid imbalance common to both
formats. Overlapping that exposed reduction with compute is a remaining lever.

\subsection{Cold setup}
\label{sec:cold}

The one-time cold setup is dominated by symbolic coarsening and prolongator smoothing. The
native-block setup steps (operator inspection, graph, $P_{\mathrm{oth}}$) perform no \texttt{AIJ}
expansion on the coarsening path---confirmed by per-stage logging showing zero conversions in the hot
second setup. The prolongator smoother's \texttt{MatAXPY} (\texttt{MatAYPX} with
\texttt{SUBSET\_NONZERO\_PATTERN}) likewise runs natively in blocks: the diagonal block is added
directly, and the off-diagonal block translates the two operands' differing off-process column
compressions through their sorted ghost-column maps and adds block-for-block on device, with no
scalar \texttt{AIJ} round trip. This removed the last residual conversion in the cold path: on a
$128^3$ grid (four A100s, two ranks per GPU) the 18 \texttt{MatConvert} calls and the $\sim6$\,s they
cost---previously $\sim6$\,s of a 13.7\,s cold setup---are eliminated; \texttt{MatConvert} is now a
negligible fraction of the cold setup and the smoother performs native block \texttt{AXPY} flops on
device, cutting the cold setup to $\approx7.9$\,s. With this the cold setup is as conversion-free as
the recurring path.

\section{Blocked COO as a Reusable Primitive}
\label{sec:coo}

The coarse-operator assembly of Section~\ref{sec:coo-design} is built on a primitive of independent
value. PETSc's COO assembly interface lets a caller declare, once, the $(i, j)$ coordinates of every
contribution to a matrix---including duplicates and entries destined for other ranks---after which
PETSc constructs a cached communication-and-scatter plan and each subsequent numeric assembly is a
single device kernel. This is the standard high-performance device-assembly path in PETSc, used most
heavily by finite-element codes, where per-element dense matrices generate exactly such a stream of
duplicated, possibly-off-process contributions.

Our \texttt{MatCOOUseBlockIndices} option generalizes this from scalar entries to dense $bs_r \times
bs_c$ value blocks. Each declared coordinate addresses a block, the value stream is a sequence of
dense blocks (square or rectangular), and the scatter writes and sums contiguous $bs_r \cdot bs_c$
runs. Everything that the scalar COO plan stores---the coordinate arrays, the duplicate-summation
map, and the off-process \texttt{PetscSF}---shrinks by a factor of about the block area. Within this
paper the primitive serves the Galerkin coarse-operator assembly. Its reach is broader: any caller
assembling a blocked matrix on the device---finite-element assembly of vector PDEs being the immediate
example---can emit block coordinates and a block value stream and obtain the same block-area reduction
in index overhead and assembly traffic. The primitive is tested for sequential and parallel blocked
matrices. A standalone characterization of blocked COO against element-by-element assembly for a
finite-element workload is left to future work.

\section{Conclusion and Future Work}
\label{sec:conclusion}

We have made PETSc's smoothed-aggregation algebraic multigrid run in a blocked format with no scalar
\texttt{AIJ} expansion on the coarsening path, and with its two recurring phases---the Galerkin
coarse-operator recompute (\PtAP) and the V-cycle---resident on the GPU in blocks. Each setup step
(operator inspection, the aggregation graph, the off-process prolongator gather, and coarse-operator
assembly) operates on the block format directly, and cached, state-gated reuse keeps the repeated
Galerkin recompute on the device with no host round trip. On A100 GPUs for 3D elasticity the blocked
format extends the problem size solvable with the vendor cuSPARSE backend---which runs out of memory
forming the coarse Galerkin product for a $128^3$ grid on 8 GPUs, where the blocked format and the
portable native-Kokkos-Kernels scalar path both fit---and, where the formats run, is faster at scale
($1.04\times/1.24\times/1.16\times$ \KSP, $1.12\times/1.42\times/1.30\times$ SpMV,
$1.45\times/1.80\times/2.27\times$ hot \PtAP at 8/27/64 GPUs) and at parity on one GPU. The \PtAP
advantage grows monotonically with GPU count; the SpMV and \KSP advantages peak at 27 GPUs, where the
SpMV ratio reaches the $1.42\times$ index-bandwidth ceiling, and narrow at 64 GPUs as the scalar
cuSPARSE path scales more efficiently. The advantage is bandwidth-bound in origin: it is largest in
the low-nonzeros-per-row regime, where index compression matters most, and narrows as the kernels
become flop-bound.

The blocked COO assembly primitive (\texttt{MatCOOUseBlockIndices}) is a contribution beyond GAMG:
it brings the same $bs^2$ index-and-traffic reduction to any device-resident blocked assembly, with
finite-element assembly the natural next application.

\textbf{Device-resident MIS coarsening} (the aggregation graph phase of Section~\ref{sec:graph})
completes the cold setup. A new Kokkos coarsener (\texttt{MATCOARSENMISKOKKOS}) runs parallel
Luby-round MIS on the device with deterministic hash weights and is integrated into GAMG's
auto-selection path for device-resident operators. The MPI case---where the boundary-vertex
neighborhood spans ranks---is handled by a device-resident \texttt{PetscSF} ghost-layer exchange of
the per-vertex aggregate-owner array, reusing the same memtype-aware communication path as the
blocked SpMV halo so the owner array never leaves device memory between Luby rounds; the hash-weight
total order makes the selection independent of the rank count. Aggressive coarsening follows without
additional kernels: GAMG's default aggressive strategy squares the graph, and the squared
\texttt{MPIAIJKOKKOS} operator routes through the same device coarsener. On A100 GPUs the device
coarsener produces aggregates that match the CPU MIS coarsener and yields identical parallel solver
convergence, so the aggregation graph phase now runs entirely on the GPU.

Several directions remain. \textbf{Overlapping the off-process reduction} in the hot \PtAP with
compute would attack the exposed, imbalance-gated wait that now dominates that phase at scale
(Section~\ref{sec:comm}). \textbf{Vendor-specific blocked kernels} are a further direction: the
portable Kokkos \MATBAIJ kernels developed here serve as a reference implementation and template
from which hand-tuned vendor versions---CUDA for NVIDIA and HIP for AMD---can be derived, and, as
the vendor sparse libraries mature, the rectangular-block kernels could eventually migrate onto
cuSPARSE and rocSPARSE (the latter's GEBSR format already exposing the independent row and column
block sizes that smoothed-aggregation elasticity requires). Finally, a \textbf{standalone evaluation
of blocked COO} for finite-element assembly would quantify the primitive's benefit in its own right.

\section*{Code Availability}

The implementation described in this paper is part of PETSc~\cite{PETSc} and is available in the
PETSc git repository at \url{https://gitlab.com/petsc/petsc} on the branch
\texttt{adams/\allowbreak seqbaijkokkos-\allowbreak gamg}. The model problems are the elasticity
examples \texttt{src/ksp/ksp/\allowbreak tutorials/ex56} and
\texttt{src/snes/\allowbreak tutorials/ex56}.

\section*{Use of AI}

Portions of this work---including code development and debugging, performance-data reduction and
analysis, and preparation of this manuscript---were carried out with the assistance of Anthropic's
Claude (Claude Code). All design decisions, experiments, and results were directed, generated, and
verified by the author, who takes full responsibility for the content.

\section*{Acknowledgment}

This material is based upon work supported by the U.S.\ Department of Energy, Office of Science,
Office of Advanced Scientific Computing Research, Scientific Discovery through Advanced Computing
(SciDAC) Program through the FASTMath Institute, under contract number DE-AC02-05CH11231 at
Lawrence Berkeley National Laboratory.

\bibliographystyle{IEEEtran}
\bibliography{refs}

\end{document}